\documentclass{jfm}
\usepackage{graphicx}
\usepackage{color}

\shorttitle{Deep learning prediction of vortex shedding over a cylinder}
\shortauthor{S. Lee and D. You}

\title{Prediction of laminar vortex shedding over a cylinder using deep learning}

\author{Sangseung Lee
  \and Donghyun You\corresp{\email{dhyou@postech.ac.kr}}}

\affiliation{Department of Mechanical Engineering, 
Pohang University of Science and Technology,\\ 77 Cheongam-ro, Nam-gu, Pohang, Gyeongbuk 37673, Republic of Korea}

\begin{document}

\maketitle

\begin{abstract}
Unsteady laminar vortex shedding over a circular cylinder is predicted using a deep learning technique, a generative adversarial network (GAN), with a particular emphasis on elucidating the potential of learning the solution of the Navier-Stokes equations. Numerical simulations at two different Reynolds numbers with different time-step sizes are conducted to produce training datasets of flow field variables. Unsteady flow fields in the future at a Reynolds number which is not in the training datasets are predicted using a GAN. Predicted flow fields are found to qualitatively and quantitatively agree well with flow fields calculated by numerical simulations. The present study suggests that a deep learning technique can be utilized for prediction of laminar wake flow in lieu of solving the Navier-Stokes equations.
\end{abstract}

\begin{keywords}
To be added during the typesetting process
\end{keywords}

\section{Introduction}\label{sec:introduction}
Observations of fluid flow in nature, laboratory experiments, and numerical simulations have provided evidence of existence of flow features and certain, but often complex, ordinance. For example, in nature, Kelvin-Helmholtz waves in the cloud~\citep{dalin2010case}, von Karman vortices in ocean flow around an island~\citep{berger1972periodic}, and swirling great red spot on the Jupiter~\citep{marcus1988numerical} are flow structures that can be classified as a certain type of vortical motions produced by distinct combination of boundary conditions and initial conditions for the governing first principles. Similar observations also have been reported in laboratory experiments and numerical simulations~\citep{freymuth1966transition,ruderich1986experimental,wu2009direct,babucke2008dns}. The existence of distinct and dominant flow features that are linearly independent have been also widely investigated by mathematical decomposition techniques such as the proper orthogonal decomposition (POD) method~\citep{sirovich1987turbulence} or the Koopman operator method~\citep{mezic2013analysis}.

For the sake of the existence of distinct or dominant flow features, animals such as insects, birds, and fish are reported to control their bodies adequately to better adapt the fluid dynamic environment and to improve the aero- or hydro-dynamic performance and efficiency~\citep{wu2011fish}. This suggests a possibility that they empirically learn dominant fluid motions as well as the non-linear correlation of fluid motions and are able to estimate the future flow based on the experienced flow in their living environments. 
However, it is highly unlikely that the learning procedures are based on numerical simulations or mathematical decomposition techniques. 
Such observation in nature motivates us to investigate the feasibility of predicting unsteady fluid motions by learning flow features using a deep learning technique.

An attempt to apply deep learning on fluid flow has been recently conducted by~\citet{ling2016reynolds} using a deep neural network (DNN) to model the Reynolds stress anisotropy tensor for a Reynolds-average Navier-Stokes (RANS) simulation. This approach was found to notably improve the accuracy of the simulation result. \citet{guo2016convolutional} employed a convolutional neural network (CNN) to predict steady flow fields around bluff objects and reported reasonable prediction of steady flow fields with significantly reduced computational cost than that required for numerical simulations. Similarly, \citet{miyanawala2017efficient} employed a CNN to predict aerodynamic force coefficients of bluff bodies, also with notably reduced computational costs. 
Those previous studies showed high potential of deep learning techniques for enhancing simulation accuracy and reducing computational costs. 
However, to the best of our knowledge, no earlier research to directly predict unsteady flow fields using a deep learning technique has been reported.

Prediction of unsteady flow fields using deep learning could offer new opportunities for real-time control and guidance of aero- or hydro-vehichles, fast weather forecast, {\it etc.}, due to the significantly low computational cost compared to numerical simulations of the Navier-Stokes equations. As the first step towards prediction of unsteady flow fields using deep learning, in the present study, it is attempted to predict rather simple but canonical unsteady laminar vortex shedding over a circular cylinder. A generative adversarial network (GAN) with a multi-scale convolution architecture proposed by~\citet{mathieu2015deep} is employed as a deep learning method.

The paper is organized as follows. Numerical methods for Navier-Stokes simulations are explained in section 2. The method for construction of training datasets is explained in section 3. The present deep learning method and results obtained using the deep learning method are presented in sections 4 and 5, respectively, followed by concluding remarks in section 6.

\section{Numerical simulations}\label{sec:numerical simulations}
\begin{figure}
\centerline{\includegraphics[width = 0.8 \linewidth,trim={0 2cm 0 0cm},clip]{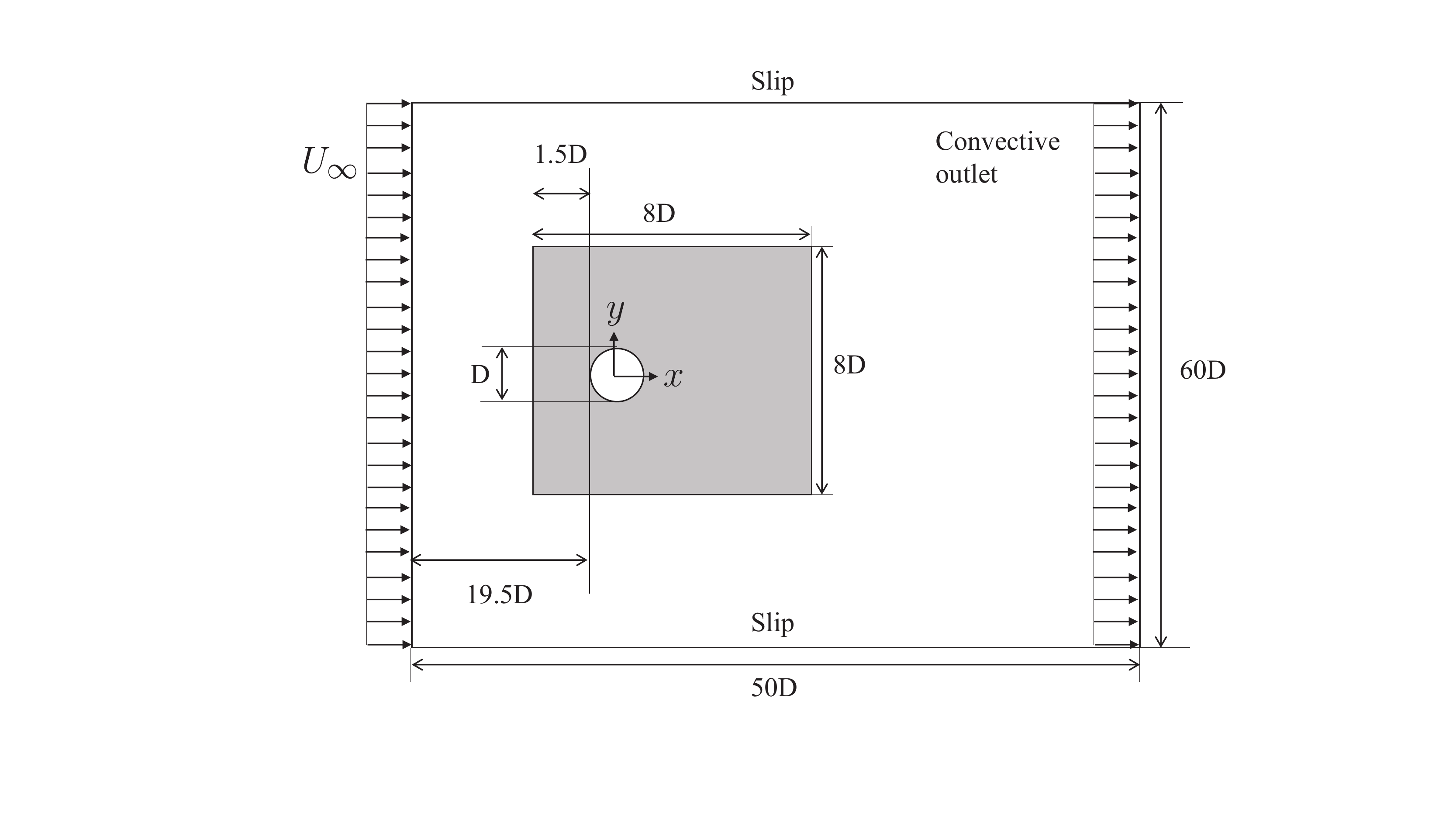}}
\caption{The computational domain for numerical simulations (whole domain) with boundary conditions and the training domain for collecting datasets for the DNN (gray area).}
\label{fig: com_domain} 
\end{figure}

Numerical simulations of flow over a circular cylinder at $\Rey_{D} = U_\infty D / \nu = 100, 160$, and $300$, where $U_\infty$, $D$, and $\nu$ are the freestream velocity, cylinder diameter, and kinematic viscosity, respectively, have been conducted by solving the incompressible Navier-Stokes equations as follows:
\begin{eqnarray}
\frac{\partial u_{i}}{\partial t} + \frac{\partial u_{i}u_{j}}{\partial u_{j}} = -\frac{1}{\rho}\frac{\partial p}{\partial x_{i}} + \nu \frac{\partial^2 u_{i}}{\partial x_j \partial x_{j}}
\label{eq_momentum_diff}
\end{eqnarray}
and
\begin{eqnarray}
\frac{\partial u_{i}}{\partial x_{i}} = 0,
\label{eq_continuity_diff}
\end{eqnarray}
where $u_{i}$, $p$, and $\rho$ are the velocity, pressure, and density, respectively. A fully implicit fractional step method is employed as a time integration technique, whereas all terms in the Navier-Stokes equations are integrated using the Crank-Nicolson method. Second-order central-difference schemes are employed for spatial discretization~\citep{you2008discrete}. Figure~\ref{fig: com_domain} shows the computational domain with boundary conditions. The computational domain sizes are $50D$ and $60D$ in the streamwise and cross-stream directions, respectively, while the domain is discretized with $400 \times 250$ structured cells. The computational time-step sizes ($\Delta t U_{\infty} / D$) of  $0.003$, $0.0048$, and $0.009$ are used for simulations at $\Rey_{D}$ of 100, 160, and 300, respectively. The domain size, number of cells, and time-step sizes are determined from an extensive sensitivity study.

\section{Training datasets}
\begin{figure}
\centerline{\includegraphics[width = 0.5 \linewidth,trim={0 0cm 0 0cm},clip]{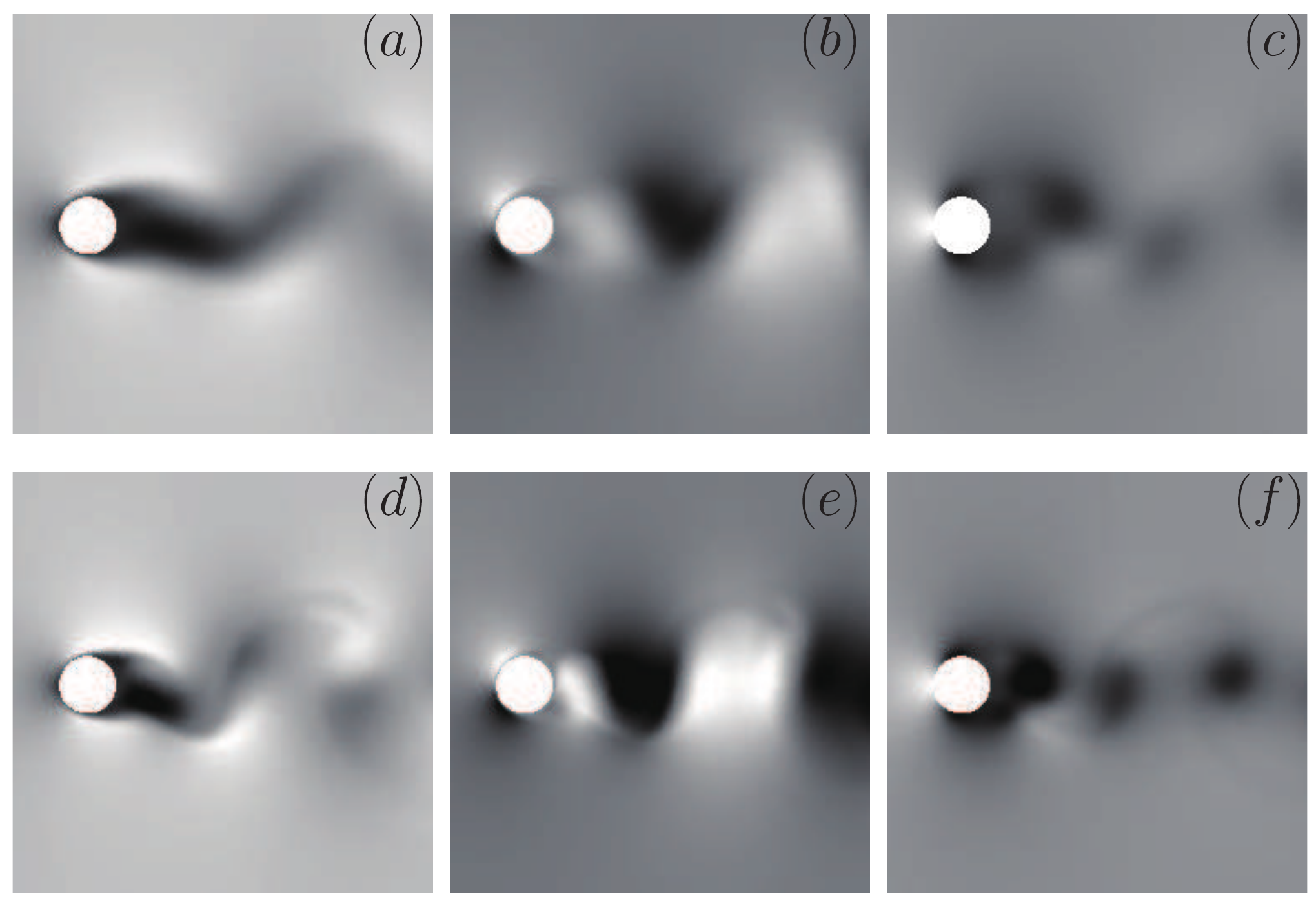}}
\caption{Snapshots of flow fields in the training dataset. (a) $u/U_{\infty}$, (b) $v/U_{\infty}$, (c) $p/\frac{1}{2} \rho U_{\infty}^{2}$ at $\Rey_{D}=100$ and (d) $u/U_{\infty}$, (e) $v/U_{\infty}$, (f) $p/\frac{1}{2} \rho U_{\infty}^{2}$ at $\Rey_{D}=300$. $u/U_{\infty}$ is ranged from -0.3 (black) to 1.4 (white); $v/U_{\infty}$ is ranged from -0.7 (black) to 0.7 (white); $p/\frac{1}{2} \rho U_{\infty}^{2}$ is ranged from -1.4 (black) to 1.0 (white).}
\label{fig: gans_training_data} 
\end{figure}

Videos of flow variables ($u_1 (=u)$, $u_2 (=v)$, and $p$) in the training domain (gray area, $-2< x/D < 6$ and $-4< y/D < 4$) as shown in figure~\ref{fig: com_domain} are randomly cropped in space and time into five million clips as the training dataset.
Each clip again contains five consecutive images of a flow variable, where the first four consecutive images are used as inputs and the last image is used as the ground truth image, which is expected to be predicted. 
The flow field data is fed into the deep learning routine, which will be discussed in the next section, as images in the format of png with three feature maps (R(red)G(reen)B(lue)), but colored in a gray scale.
The time-step intervals of the images are $\Delta t U_{\infty}/D = 0.06$ and 0.18 for flow at $\Rey_{D}=100$ and $300$, respectively.
Snapshots of flow fields in the training dataset are shown in figure~\ref{fig: gans_training_data}. 
In the present study, the network is first trained only for solutions at $Re_D$ of 100 and 300, and then tested for prediction of solutions at various future occasions for $Re_D$ of 160.

\section{Deep learning methodology}\label{sec:predictions}
\subsection{Architecture of the generative adversarial network}
\begin{figure}
\centerline{\includegraphics[width = 0.65 \linewidth,trim={0 0cm 0 0cm},clip]{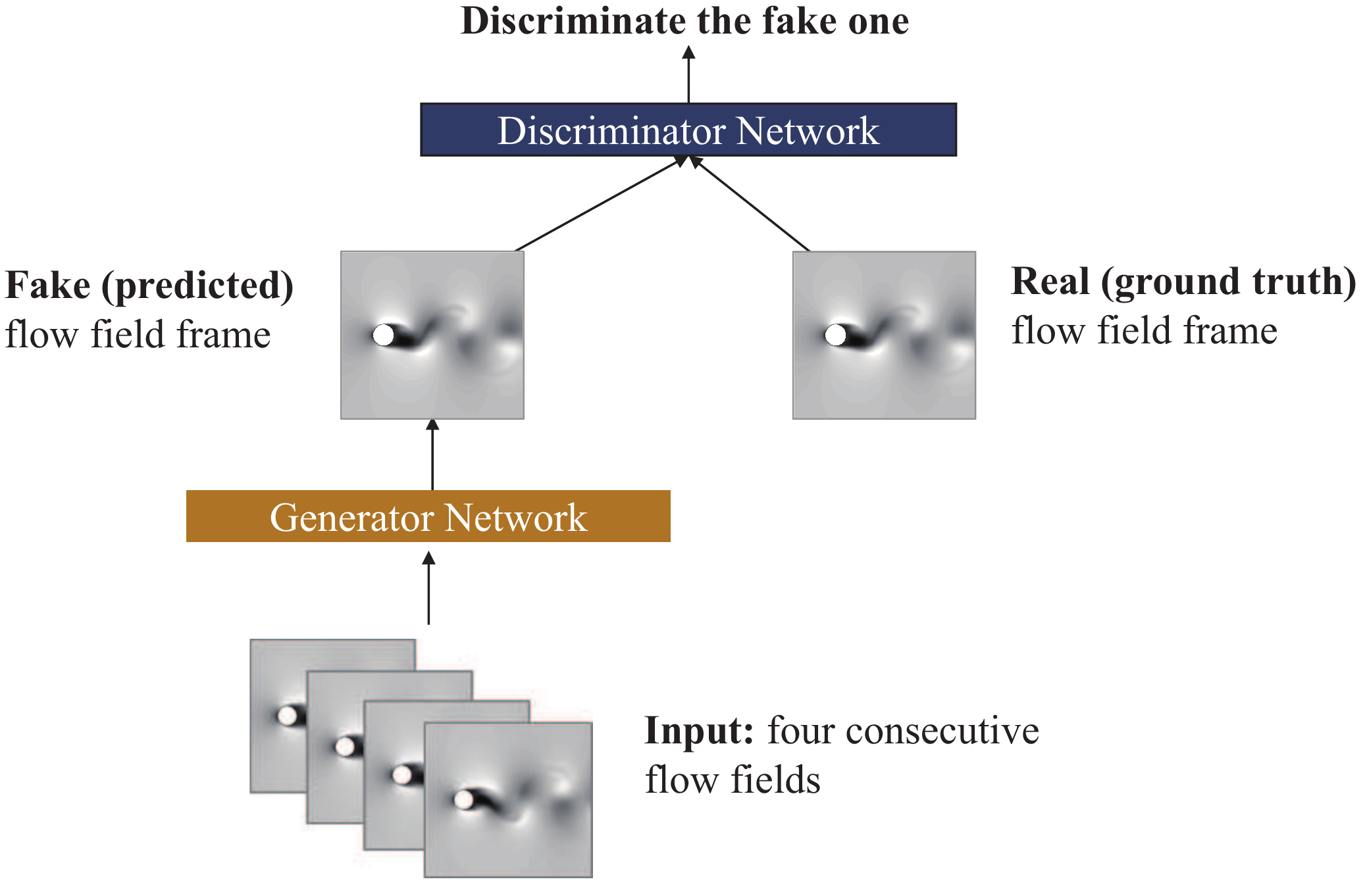}}
\caption{Schematic diagram of the overall architecture of the implemented GAN.}
\label{fig: gans_arch} 
\end{figure}

A GAN consists of a generator network and a discriminator network. The generator network generates fake images and the discriminator network is employed to discriminate the generated fake images from real (ground truth) images. The general goal of training a GAN is to fool the discriminator by generating a real-world like image. In the present study, it is to generate a well predicted flow field image that can fool the discriminator. The overall architecture of the present GAN is shown in figure~\ref{fig: gans_arch}. The GAN architecture and the loss function proposed by~\citet{mathieu2015deep} are utilized in the present study to predict future flow fields.
 
\begin{figure}
\centerline{\includegraphics[width = 0.7 \linewidth,trim={0 0cm 0 0cm},clip]{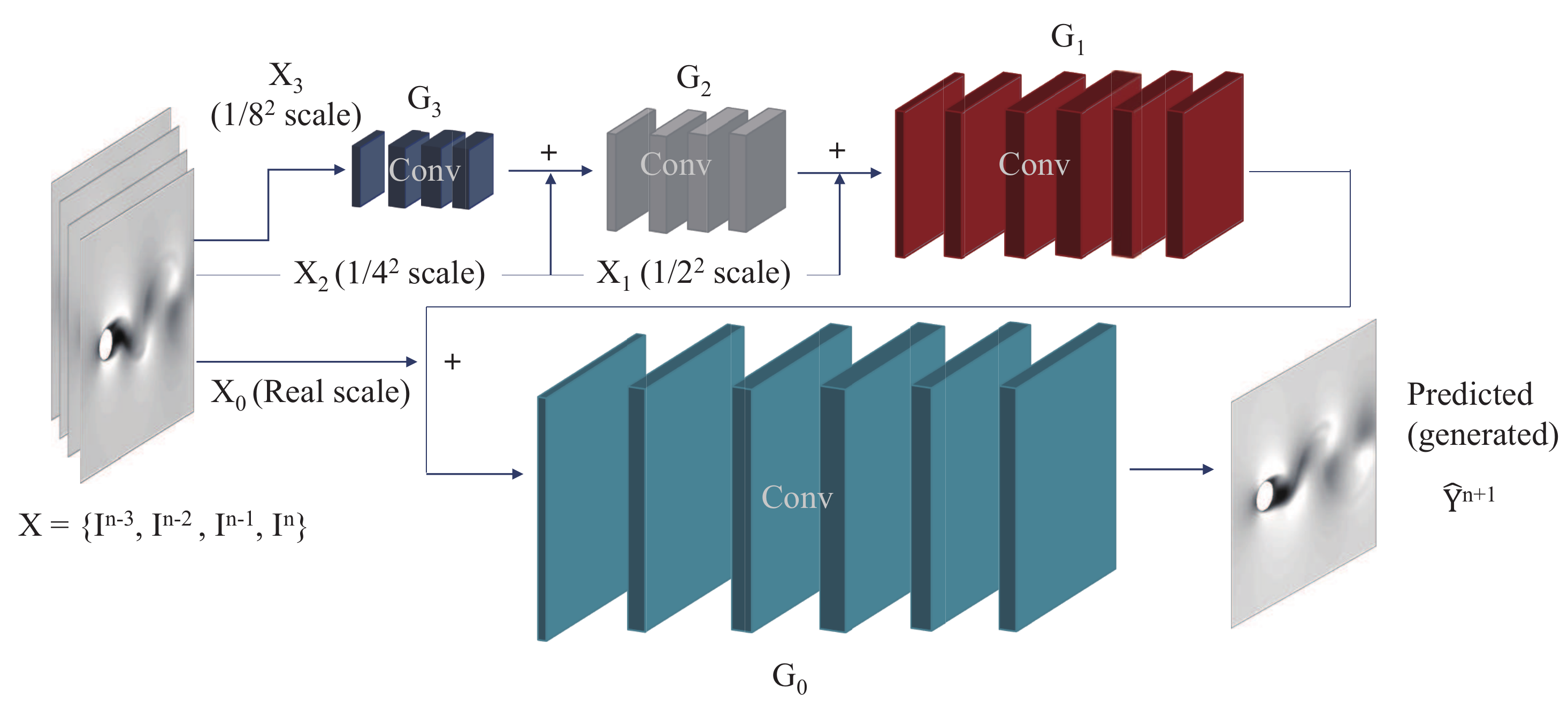}}
\caption{Schematic diagram of the generator network. Numbers of feature maps of the multi-scale convolutional neural networks are $3 \times 4$, 128, 256, 128, 3 on the $1/8^{2}$ scale;  $3 \times 5$, 128, 256, 128, 3 on the $1/4^{2}$ scale; $3 \times 5$, 128, 256, 512, 256, 128, 3 on the $1/2^{2}$ scale; $3 \times 5$, 128, 256, 512, 256, 128, 3 on the real scale. The convolution filter sizes are $3\times3$, $3\times3$, $3\times3$, $3\times3$ on the $1/8^{2}$ scale; $5\times5$, $3\times3$, $3\times3$, $5\times5$ on the $1/4^{2}$ scale; $5\times5$, $3\times3$, $3\times3$, $3\times3$, $3\times3$, $5\times5$ on the $1/2^{2}$ scale; $7\times7$, $5\times5$, $5\times5$, $5\times5$, $5\times5$, $7\times7$ on the real scale.}
\label{fig: generator} 
\end{figure}

A set of four consecutive flow field images $X = \{I^{n-3}, I^{n-2}, I^{n-1}, I^{n}\}$, where $I$ denotes the input image and the superscript denotes the time step sequence of the image, are fed as input images to the generator network $G$, which is composed of fully convolutional layers with multi-scale architectures $\{G_{0}, G_{1}, G_{2}, G_{3}\}$ (see figure~\ref{fig: generator}). The set of input images ($X=X_{0}$) is rescaled into $1/2^{2}$ ($X_{1}$), $1/4^{2}$ ($X_{2}$), and $1/8^{2}$ ($X_{3}$) scales to reduce the loss of resolution due to small sizes of convolution filters. The generator network at scale $1/(2^{k})^{2}$ generates a predicted flow field at the next time step $\hat{Y}_{k}^{n+1}$ based on an upscaled image from the one lower level resolution network and a rescaled set of input images as
\begin{equation}
   \hat{Y}_{k}^{n+1} = \left\{
    \begin{array}{ll}
     G_{k}\left(X_{k}\right), & {\hbox{for}} \ k = 3, \\[2pt]
     G_{k}(X_{k} , u_{k}(\hat{Y}_{k+1}^{n+1})), & otherwise,
    \end{array} \right.
\end{equation}
where $u_{k}$ is the one level upscaling operator from scale $1/(2^{k+1})^{2}$ to $1/(2^{k})^{2}$ and $X_{k}$ is the rescaled set of input images for $k=0,1,2$ and $3$.
Based on these generator architecture, the generator network tries to generate a well predicted flow field $\hat{Y}^{n+1}  = \hat{Y}_{0}^{n+1}$.

\begin{figure}
\centerline{\includegraphics[width = 0.6 \linewidth,trim={0 0cm 0 0cm},clip]{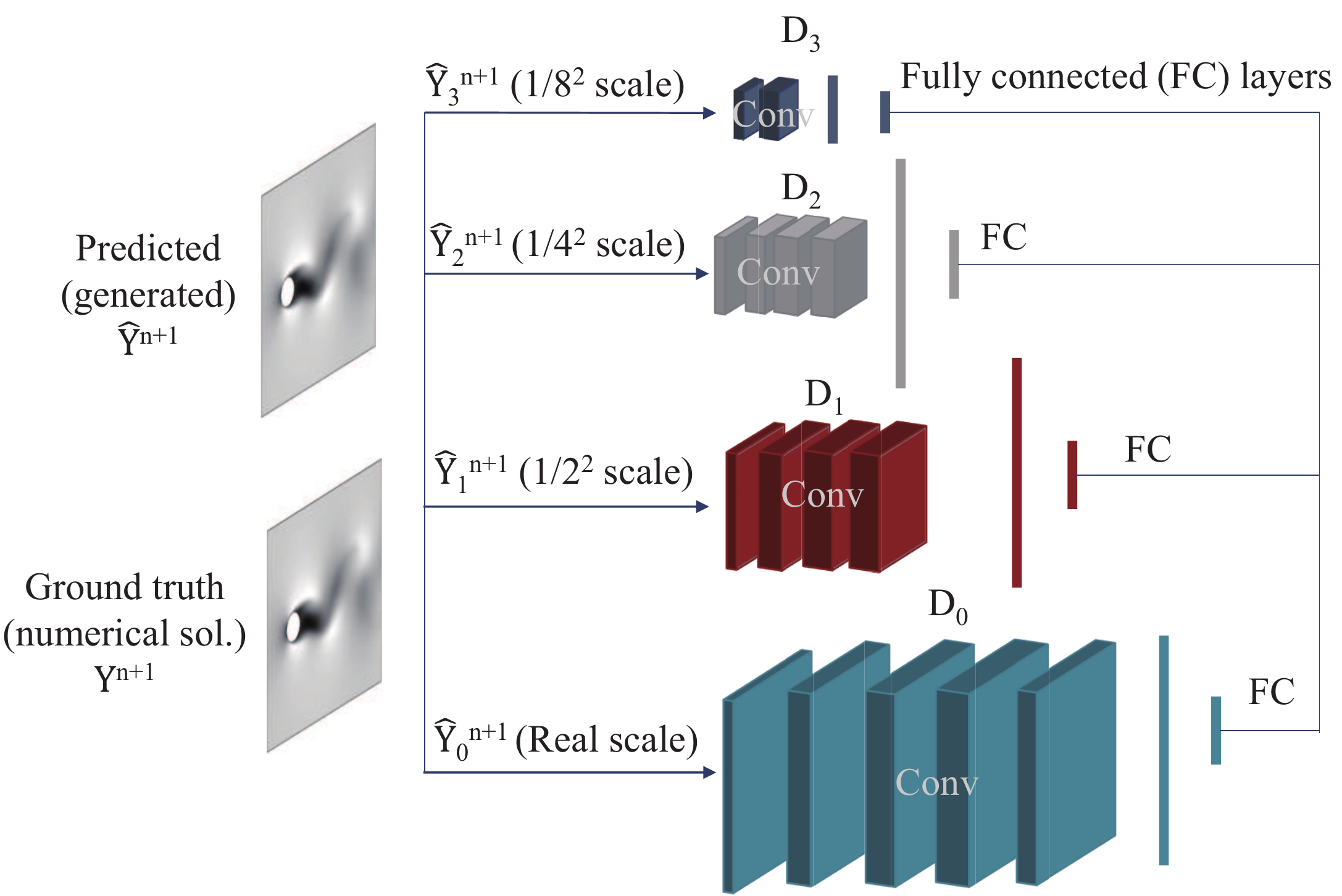}}
\caption{Schematic diagram of the discriminator network. Numbers of feature maps of the multi-scale convolutional neural networks are 3, 64 on the $1/8^{2}$ scale; 3, 64, 128, 128 on the $1/4^{2}$ scale; 3, 128, 256, 256 on the $1/2^{2}$ scale; 3, 128, 512, 256, 512, 128 on the real scale. The convolution filter sizes are $3\times3$ on the $1/8^{2}$ scale; $3\times3$, $3\times3$, $3\times3$ on the $1/4^{2}$ scale; $5\times5$, $5\times5$, $5\times5$ on the $1/2^{2}$ scale; $7\times7$, $7\times7$, $5\times5$, $5\times5$ on the real scale. The sizes of fully connected layers are 512, 256, 1 on the $1/8^{2}$ scale; 1024, 512, 1 on the $1/4^{2}$, $1/2^{2}$, and real scales.}
\label{fig: discriminator} 
\end{figure}

The discriminator network looks into both the generated flow field and the ground truth flow field that is obtained from a numerical simulation, and tries to discriminate the generated flow field (figure~\ref{fig: discriminator}). The discriminator network $D$ has a similar multi-scale architecture $\{D_{0}, D_{1}, D_{2}, D_{3}\}$ to the generator network, but with additional fully-connected layers for single scalar outputs in the range of 0 to 1 for each scale, where output value 1 means real and 0 means fake. Let $Y_{k}^{n+1}$ be the rescaled ground truth flow field image at $1/(2^{k})^{2}$ scale, then the discriminator network discriminates the fake flow field image at each scale by evaluating output scalar values from $D_{k}(Y_{k}^{n+1})$ and $D_{k}(\hat{Y}_{k}^{n+1})$ for $k=0,1,2$, and $3$.

\subsection{Loss function of the generative adversarial network}
For a given set of input flow field images and a ground truth flow field image $Y^{n+1}$, the generator network $G$ generates a flow field image $\hat{Y}^{n+1}$ that minimizes a loss function which is a combination of three different loss functions~\citep{mathieu2015deep} as follows:
\begin{equation}
L(\hat{Y}^{n+1},Y^{n+1}) =  L_{2}(\hat{Y}^{n+1},Y^{n+1}) + L_{gdl}(\hat{Y}^{n+1},Y^{n+1})  + 0.05 L_{adv}^{G}(\hat{Y}^{n+1}).
\label{eqn: L_gen}
\end{equation}
$L_{2}(\hat{Y}^{n+1},Y^{n+1})$ is the square of the $L_{2}$ error loss function which tries to minimize the difference between the generated flow field image and the ground truth flow field image as
\begin{equation}
L_{2}(\hat{Y}^{n+1},Y^{n+1}) = \frac{1}{N}\sum_{k=0}^{N-1} ||\hat{Y}_{k}^{n+1} - Y_{k}^{n+1}||_{2}^{2},
\label{eqn: ll2}
\end{equation}
where $N$ is the number of scales.
However, the $L_{2}$ error loss function inherently induces a blurry prediction. Therefore, a gradient difference loss (GDL) function is introduced to sharpen a flow field image by directly penalizing the gradients of the generated flow field, which prevents the predicted flow field to be overly diffused, as follows
\begin{eqnarray}
L_{gdl}(\hat{Y}^{n+1},Y^{n+1}) = \frac{1}{N}\sum_{k=0}^{N-1}\sum_{i,j}||Y_{k,(i,j)}^{n+1}-Y_{k,(i-1,j)}^{n+1}|-|\hat{Y}_{k,(i,j)}^{n+1} - \hat{Y}_{k,(i-1,j)}^{n+1}|| + \nonumber\\
||Y_{k,(i,j-1)}^{n+1}-Y_{k,(i,j)}^{n+1}|  -|\hat{Y}_{k,(i,j-1)}^{n+1} - \hat{Y}_{k,(i,j)}^{n+1}||,
\label{eqn: lgdl}
\end{eqnarray}
where the subscript $(i,j)$ indicates the pixel coordinate of a flow field image.

$L_{adv}^{G}(\hat{Y}^{n+1})$ is a loss function which tries to fool the discriminator network by generating a well predicted flow field image at each scale, so it minimizes
\begin{equation}
L_{adv}^{G}(\hat{Y}^{n+1}) = \frac{1}{N}\sum_{k=0}^{N-1} L_{bce}(D_{k}(\hat{Y}_{k}^{n+1}),1),
\label{eqn: ladvg}
\end{equation}
where $L_{bce}$ is the binary cross entropy loss function defined as
\begin{equation}
L_{bce}(a,b) = -b \log(a) - (1-b) \log(1-a),
\label{eqn: lbce}
\end{equation}
for scalar values a and b.

The discriminator network tries to judge the ground truth flow field image at each scale as real and the generated flow field image at each scale as fake, thus it minimizes
\begin{equation}
L_{adv}^{D}(\hat{Y}^{n+1},Y^{n+1}) = \frac{1}{N}\sum_{k=0}^{N-1} L_{bce}(D_{k}(Y_{k}^{n+1}),1) + L_{bce}(D_{k}(\hat{Y}_{k}^{n+1}),0).
\label{eqn: ladvd}
\end{equation}

The generator network is trained with an Adam optimizer~\citep{adam14} with the learning rate, which is a parameter that determines the network update speed, of $4 \times 10^{-5}$ and the discriminator network is trained with the gradient descent method with the learning rate of $0.02$.

\section{Results and discussion}
\subsection{Flow field prediction based on ground truth inputs}
A flow field for each flow variable at $\Rey_{D}=160$, at which Reynolds number the network has not been trained, is predicted after a single time step from ground truth flow field images at four consecutive previous time-steps $\{Y^{n-3}, Y^{n-2}, Y^{n-1}, Y^{n}\}$, which are obtained from the numerical simulation. Images of flow field variables are predicted with time-step sizes of $\Delta t U_{\infty}/D = 0.096$, $0.192$, $0.288$, $0.384$, and $0.480$. Note that time-step sizes for training are $\Delta t U_{\infty}/D = 0.06$ for $\Rey_{D}=100$ and  $\Delta t U_{\infty}/D = 0.18$ for $\Rey_{D} = 300$, and the simulation time-step size is $\Delta t U_{\infty}/D = 0.0048$ for $\Rey_{D}=160$. Therefore, the maximum time-step size for prediction (0.480) is 100 times larger than the time-step size for the numerical simulation.

\begin{figure}
\centerline{\includegraphics[width = 0.45 \linewidth,trim={0 0cm 0 0cm},clip]{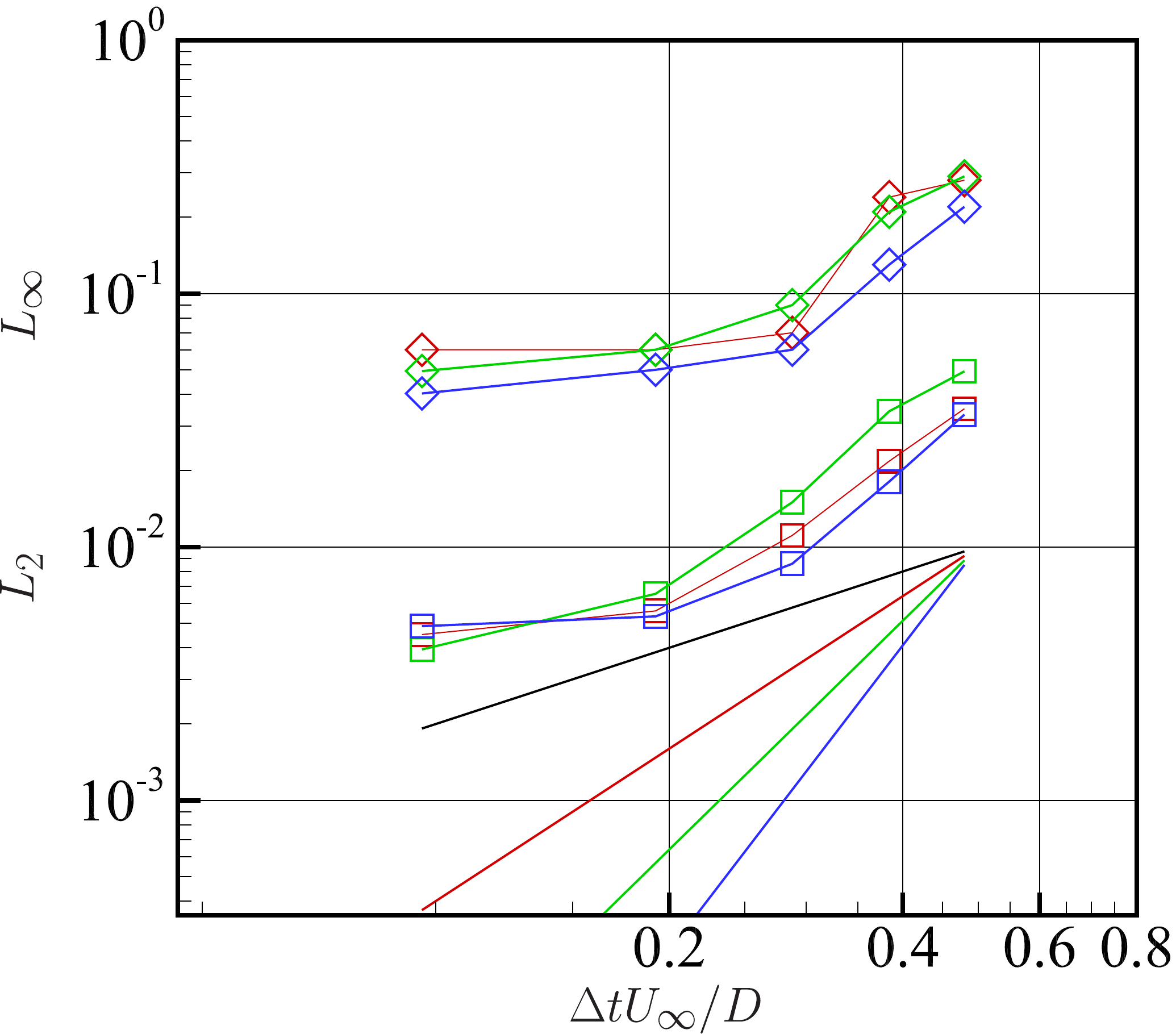}}
\caption{$L_2$ and $L_\infty$ errors for each flow field variable predicted by the GAN after a single time step using  ground truth inputs as a function of time-step sizes of $\Delta t U_{\infty}/D = 0.096$, $0.192$, $0.288$, $0.384$, and $0.480$. $L_{2}$ errors for $u/U_{\infty}$, $v/U_{\infty}$, and $p/\frac{1}{2}\rho U_{\infty}^{2}$ are denoted by \textcolor{red}{$-\square-$}, \textcolor{green}{$-\square-$}, and \textcolor{blue}{$-\square-$}, respectively; $L_{\infty}$ errors for $u/U_{\infty}$, $v/U_{\infty}$, and $p/\frac{1}{2}\rho U_{\infty}^{2}$ are denoted by \textcolor{red}{$-\diamond-$},  \textcolor{green}{$-\diamond-$}, and \textcolor{blue}{$-\diamond-$}, respectively.  \textbf{\textcolor{black}{---}}, \textbf{\textcolor{red}{---}}, \textbf{\textcolor{green}{---}}, and \textbf{\textcolor{blue}{---}} denote the 1st, 2nd, 3rd, and 4th order slopes.}
\label{fig: acc_1} 
\end{figure}

$L_{2}$ and $L_{\infty}$ errors are evaluated for the predicted flow field compared to ground truth inputs in figure~\ref{fig: acc_1}. The $L_{2}$ errors are in the order of $10^{-3}$ to $10^{-2}$, while the $L_{\infty}$ errors are in the order of $10^{-2}$ to $10^{-1}$. Errors are found to be concentrated in the separating shear layers.
Both the $L_{2}$ and $L_{\infty}$ errors are found to increase as the time interval $\Delta t U_{\infty}/D$ for prediction is increased. However, it is worth noting that the predicted flow field by the GAN agrees reasonably well with the flow field obtained from the numerical simulation even for time-step sizes, which are not only larger than the simulation time-step size but also the time-step sizes for training.

\begin{figure}
\centerline{\includegraphics[width = 0.7 \linewidth,trim={0 0cm 0 0cm},clip]{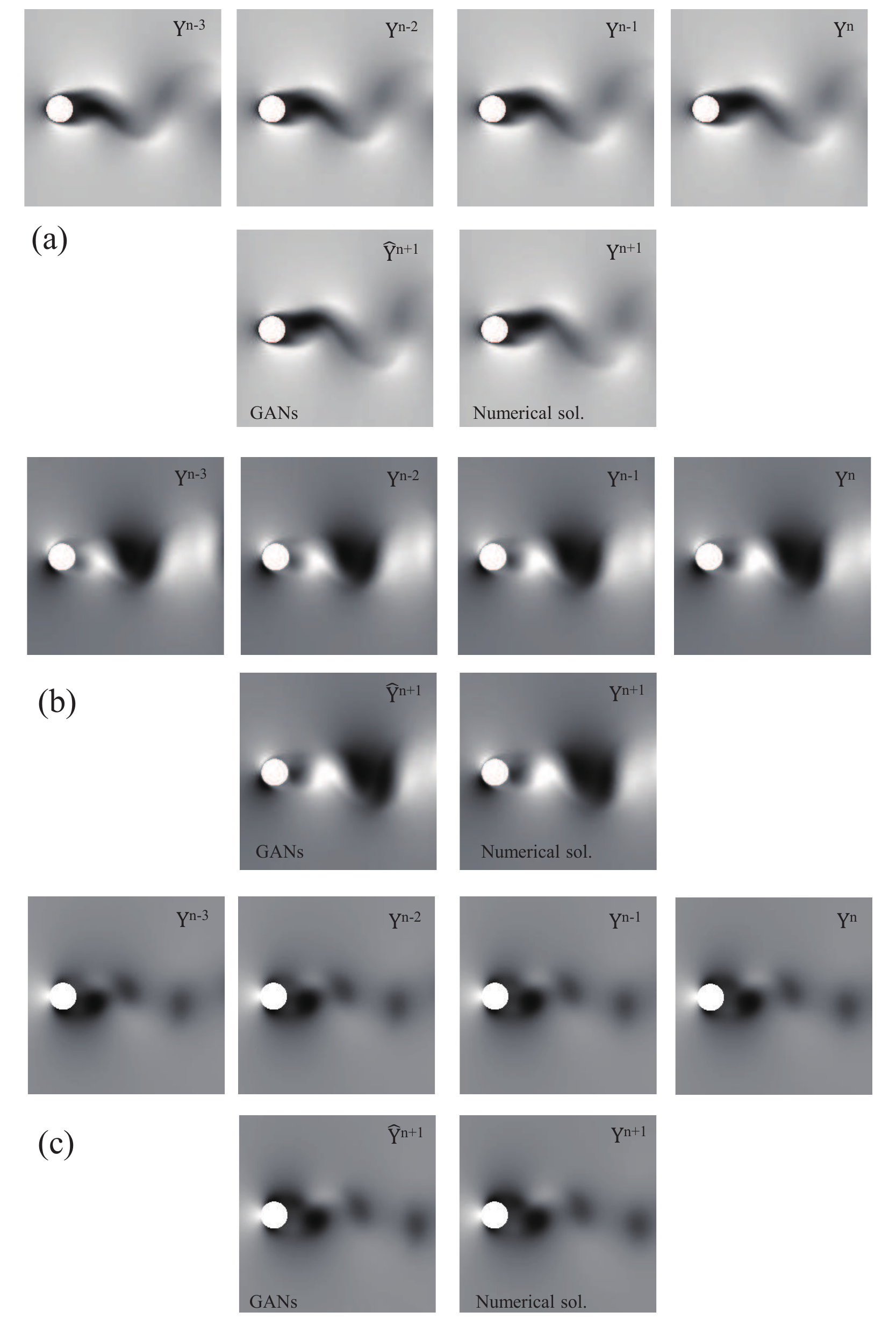}}
\caption{Comparison of flow field variables predicted by the GAN and ground truth flow fields obtained from a numerical simulation with time-step size of $\Delta t U_{\infty}/D = 0.288$. Contours of (a) $u/U_{\infty}$; (b) $v/U_{\infty}$; and (c) $p/\frac{1}{2} \rho U_{\infty}^{2}$ are shown. Contour levels for $u/U_{\infty}$, $v/U_{\infty}$, and $p/\frac{1}{2} \rho U_{\infty}^{2}$ are ranged from -0.3 (black) to 1.4 (white), from -0.7 (black) to 0.7 (white), and from -1.4 (black) to 1.0 (white), respectively.}
\label{fig: GANsresult} 
\end{figure}

A qualitative comparison of the predicted flow fields at $\Rey_{D} = 160$ with the time-step size of $\Delta t U_{\infty}/D = 0.288$ and the ground truth flow fields from the numerical simulation is shown in figure~\ref{fig: GANsresult}. 
The result shows that GAN successfully predicts flow fields at a different Reynolds number ($Re_D$) and different time-step sizes, yet the network have never seen the flow at $\Rey_{D}=160$ with a larger time-step size as they are not in the training dataset. 
The predicted patterns and magnitudes of flow variables behind the circular cylinder show good agreements with the ground truth flow fields for each flow variable. 
These results show that a GAN is capable of learning the spatial and temporal characteristics of the unsteady laminar vortex shedding phenomenon.

\subsection{Flow field prediction based on recursive inputs}
Deep learning prediction of a flow field using inputs, which are generated by the present GAN and updated recursively as the time-step advancement, is conducted to identify the ability of the network to predict flow fields not only based on the ground truth flow fields but also on the former predicted flow fields. A flow field at time step $n+m$ is recursively predicted based on the set of input flow field images $X = \{I^{n+m-4}, I^{n+m-3}, I^{n+m-2}, I^{n+m-1}\}$, where
\begin{equation}
  I^{n+m-j} = \left\{
    \begin{array}{ll}
     Y^{n+m-j}, & j \geq m \\[2pt]
     \hat{Y}^{n+m-j}, & otherwise,
    \end{array} \right.
\end{equation}
for each $j=1,2,3$, and $4$.

\begin{figure}
\centerline{\includegraphics[width = 0.45 \linewidth,trim={0 0cm 0 0cm},clip]{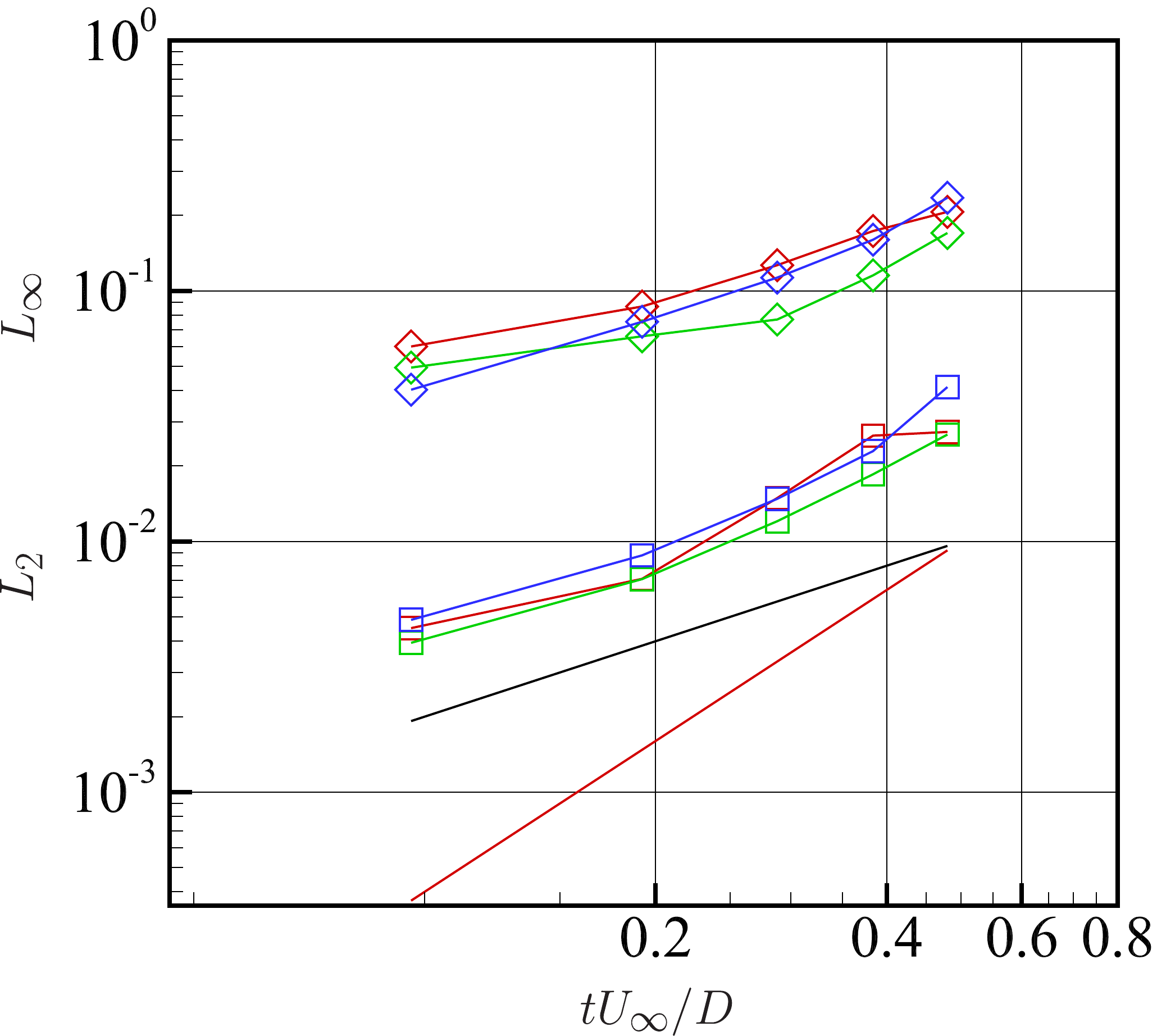}}
\caption{$L_2$ and $L_\infty$ errors for each flow field variable predicted by the GAN using recursive inputs with time-step size of $\Delta t U_{\infty}/D = 0.096$. Flow fields are predicted on time $t U_{\infty}/D = 0.096$, $0.192$, $0.288$, $0.384$, and $0.480$. $L_{2}$ errors for $u/U_{\infty}$, $v/U_{\infty}$, and $p/\frac{1}{2}\rho U_{\infty}^{2}$ are denoted by \textcolor{red}{$-\square-$}, \textcolor{green}{$-\square-$}, and \textcolor{blue}{$-\square-$}, respectively; $L_{\infty}$ errors for $u/U_{\infty}$, $v/U_{\infty}$, and $p/\frac{1}{2}\rho U_{\infty}^{2}$ are denoted by \textcolor{red}{$-\diamond-$},  \textcolor{green}{$-\diamond-$}, and \textcolor{blue}{$-\diamond-$}, respectively.  \textbf{\textcolor{black}{---}} and \textbf{\textcolor{red}{---}} denote the 1st and 2nd order slopes.}
\label{fig: acc_rec} 
\end{figure}

$L_{2}$ and $L_{\infty}$ errors are evaluated for predicted five flow field images $\hat{Y}^{n+1}$, $\hat{Y}^{n+2}$, $\hat{Y}^{n+3}$, $\hat{Y}^{n+4}$, and $\hat{Y}^{n+5}$ at $\Rey_{D}=160$ with recursive inputs, with the time-step size of $\Delta t U_{\infty}/D = 0.096$ (figure~\ref{fig: acc_rec}). Therefore, flow fields are predicted on time $t U_{\infty}/D = 0.096$, $0.192$, $0.288$, $0.384$, and $0.480$. 
$L_{2}$ errors between the recursively predicted flow fields and the ground truth flow fields are in the order of $10^{-3}$ to $10^{-2}$, while $L_{\infty}$ errors are in the order of $10^{-2}$ to $10^{-1}$. 
Likewise for results from GAN prediction based on ground truth inputs (section~5.1),  Errors are found to be concentrated in the separating shear layers.
It is worth noting that, in the prediction using recursive inputs, both $L_2$ and $L_\infty$ errors increase following the 1st and 2nd order slopes, while, in the prediction using only ground truth inputs, errors are found to increase rather rapidly (see figure~\ref{fig: acc_1}), even though in both cases, it is attempted to predict flow field variables at the same future occasion. In the present study, it is found that predicting flow fields at far future using recursive inputs with a small time-step size shows a better accuracy than predicting flow fields using only ground truth inputs with a single large time-step size.

\begin{figure}
\centerline{\includegraphics[width = 0.7 \linewidth,trim={0 0cm 0 0cm},clip]{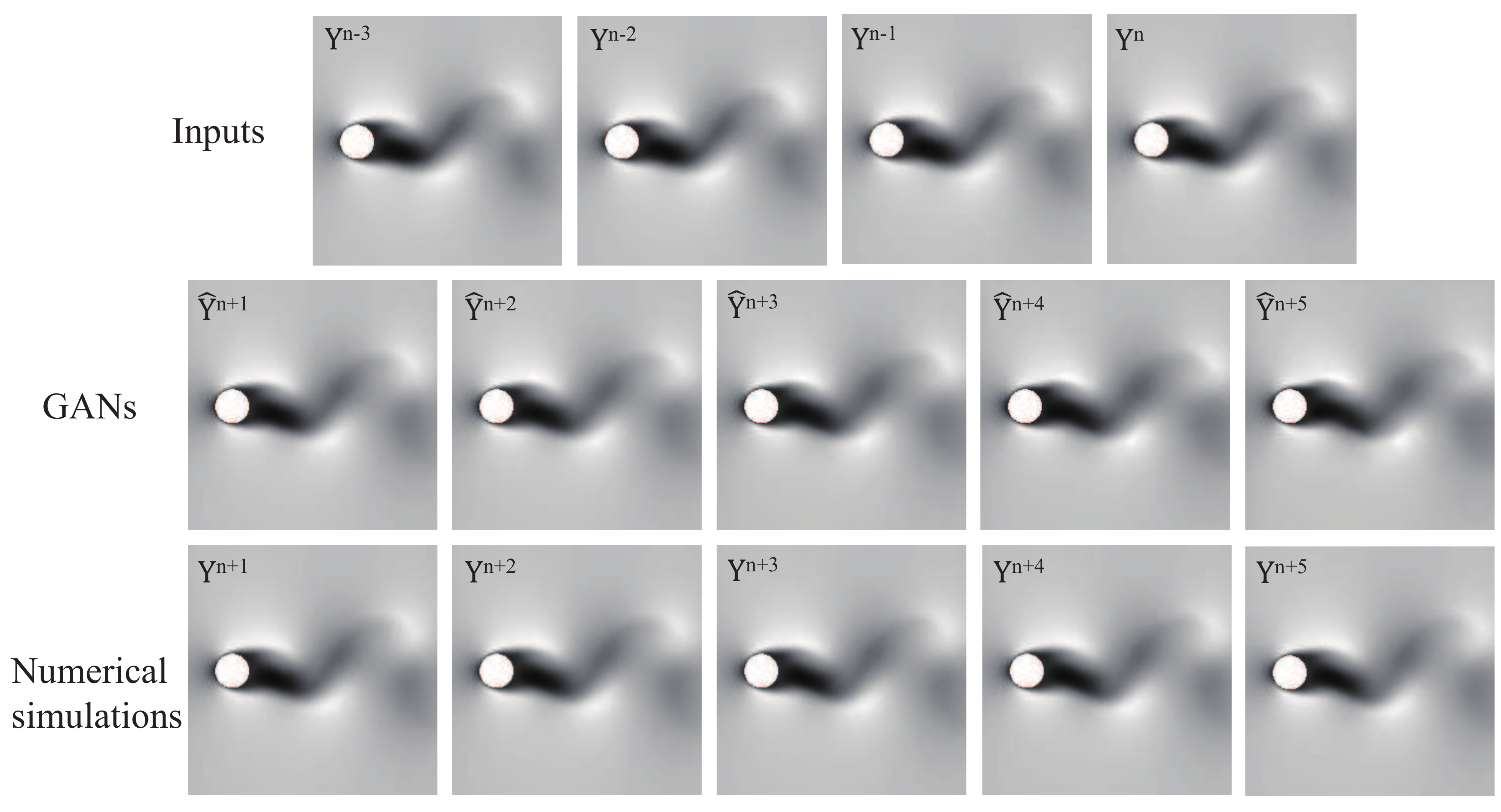}}
\caption{Comparison of recursively predicted velocity fields ($u/U_{\infty}$) by the GAN and ground truth flow fields from a numerical simulation with time-step size of $\Delta t U_{\infty}/D = 0.096$. Contour levels for $u/U_{\infty}$ are ranged from -0.3 (black) to 1.4 (white).}
\label{fig: GANsresult_rec} 
\end{figure}

A qualitative comparison of the recursively predicted flow fields of $u/U_{\infty}$ at $\Rey_{D} = 160$ with the time-step size of $\Delta t U_{\infty}/D = 0.096$ and ground truth flow fields is shown in figure~\ref{fig: GANsresult_rec}. Predicted flow fields at $\Rey_{D}=160$ show the Karman vortex wake similar to wakes at $\Rey_{D}=100$ and $300$ (see figure~\ref{fig: gans_training_data}), but with faster oscillations than the wake at $\Rey_{D}=100$ and slower oscillations than the wake at $\Rey_{D}=300$. 
The present results show that a GAN using recursive inputs is also capable of learning the spatial and temporal  characteristics of the unsteady laminar vortex shedding phenomenon.
 
\section{Concluding remarks}\label{sec:Conclusions}
Application of a deep learning method for prediction of unsteady laminar vortex shedding behind a cylinder has been explored. 
Flow fields at future occasions at $\Rey_{D}=160$, which are not in the training dataset, were successfully predicted using a generative adversarial network (GAN), which had been trained using flow field datasets produced by numerical simulations at $\Rey_{D}=100$ and $300$. 
The GAN is found to be capable of successfully learning and predicting both spatial and temporal characteristics of the laminar vortex shedding phenomenon. 
The GAN is found to be also successful in predicting flow fields with larger time-step intervals compared to time-step intervals of the training data, which are produced by numerical simulations. 
It is found that the predictive performance of the GAN can be improved by recursively updating input data for prediction.

\section{Acknowledgements}\label{sec:Acknowledgements}
This work was supported by Samsung Research Funding Center of Samsung Electronics under Project Number SRFC-TB1703-01.

\bibliographystyle{jfm}
\bibliography{2017_SLee_JFM_Rapid}

\end{document}